\documentclass{appolb}

\usepackage{graphicx}

\usepackage{amsmath}
\usepackage{braket}
\usepackage{wrapfig}

\begin{document}
\title{
	$K_S$	semileptonic decays and test of $\mathcal{CPT}$ symmetry with the KLOE detector
}
\author{Daria Kami\'nska \\ 
on behalf of the KLOE-2 collaboration
\address{
The	Marian Smoluchowski Institute of Physics, Jagiellonian University \\
{\L}ojasiewicza 11, 30-348 Krak{\'o}w, Poland
\\ daria.kaminska@uj.edu.pl}
}
\maketitle
\begin{abstract}
 Study of semileptonic decays of neutral kaons allows to perform a test of discrete symmetries, as
 well as basic principles of the Standard Model. In this paper a general review on dependency
 between charge asymmetry constructed for semileptonic decays of short- and long-lived kaons
 and $\mathcal{CPT}$ symmetry is given.
\end{abstract}
 
\section{Introduction}
 Investigations  of the neutral kaon system, due to the system's sensitivity to a variety of discrete symmetries
 such as charge conjugation~($\mathcal{C}$), parity~($\mathcal{P}$) and time reversal~($\mathcal{T}$), 
 allow to test the $\mathcal{CPT}$ symmetry as well as basic principles of the Standard Model.
 Specifically, this paper focuses on the difference and sum of charge asymmetries for the short-lived kaon ($A_S$) 
 and the long-lived kaon ($A_L$) to search for $\mathcal{CPT}$ symmetry violation.

\section{Charge asymmetry in semileptonic decays of $K_S$ meson}
 Short- and long-lived kaon states, which are hamiltonian eigenvalues, are mixture of 
 states $K^0$ and $\bar{K^0}$~\cite{handbook_cp}:

  \begin{equation}
    \ket{ K_{L/S} } = \frac{1}{\sqrt{2(1+|\epsilon_{L/S}|^2)}  } \left(  (1+ \epsilon_{L})  \ket{K^0}
					\mp (1- \epsilon_{L/S})  \ket{\bar{K^0}} \right), \\
    \label{ks_kl}
  \end{equation}
  where the parameters $\epsilon_{L}$ and $\epsilon_{S}$ account for $\mathcal{CP}$ and
  $\mathcal{CPT}$ symmetries violation.
  These parameters can be expressed in terms of $\epsilon_{K}$ and $\delta_{K}$ describing 
  $\mathcal{CP}$ and $\mathcal{CPT}$ symmetries violation, respectively:
    \begin{equation}
       \epsilon_{L/S} = \epsilon_{K} \mp \delta_{K}.
      \label{epsilon_and_symmetries}
     \end{equation}
  In order to describe semileptonic kaon decays ($K\rightarrow \pi e \nu$), due to Eq.~\ref{ks_kl},
  only the following decay amplitudes should be taken into account:
     \begin{equation}
      \begin{aligned}
      \label{aplitudes_def}
      \bra{\pi^{-} e^{+} \nu } H_{weak}  \ket{K^{0}}              &  = \mathcal{A_+} ,   &    
      \bra{\pi^{+} e^{-} \bar{\nu} } H_{weak}  \ket{ \bar{K^{0}}} &  = \mathcal{\bar{A}_{-}},\\
      \bra{\pi^{+} e^{-} \bar{\nu} } H_{weak}  \ket{K^{0}}        &  = \mathcal{A_{-}},   
						&
      \bra{\pi^{-} e^{+} \nu } H_{weak}  \ket{ \bar{K^{0}}}       &  =  \mathcal{\bar{A}_{+}},
      \end{aligned}
     \end{equation} 
   where the $H_{weak}$ is the term of Hamiltonian corresponding to the weak interaction and 
    $\mathcal{A_+}, \mathcal{\bar{A}_{-}},  \mathcal{A_{-}}, \mathcal{\bar{A}_{+}}$  %$a, b, c,  d$
  parametrize the semileptonic decay amplitudes.
  According to the Standard Model, decay of $K^0$ (or $\bar{K^0}$) state is associated with the
  transition of the $\bar{s}$ quark into $\bar{u}$ quark (or $s$ into $u$) and emission of the
  charged boson.  
  Change of strangeness ($\Delta S$) implies the corresponding
  change of electric charge  ($\Delta Q$).
  This is so called $\Delta S = \Delta Q$ rule.
  Therefore, decay  of $K^0\rightarrow \pi^- e^+ \nu$ and $\bar{K^0} \rightarrow \pi^+ e^-
  \bar{\nu}$ are present but  $ K^0 \rightarrow \pi^+ e^-\bar{\nu}$ and $\bar{K^0} \rightarrow \pi^- e^+
  \nu$ are not. This implies that, if $\Delta S = \Delta Q$ rule is conserved then parameters $\mathcal{A_{-}}$ and 
 $ \mathcal{\bar{A}_{+}}$ vanish.

  For further consideration it is useful to introduce the following notations: 
   \begin{equation}
    \begin{aligned}
    \label{zabawne_wspolczynniki}
     x  &= \frac{\mathcal{\bar{A}_+}}{\mathcal{A}_+},   
     \ \ \ \ \ \   \bar{x} = \left( \frac{ \mathcal{A_{-}}}{\mathcal{\bar{A}_{-}}} \right)^*, 
					\ \ \ \ \ \
					y = \frac{ \mathcal{ \bar{A}_{-}^{*} } - \mathcal{A}_{+}  }{  \mathcal{ \bar{A}_{-}^{*} } +
     \mathcal{A}_{+} }, \\   %= -\frac{b}{a}, \nonumber \\
      x_{\pm}  &= \frac{ x \pm \bar{x}^{*} }{2} = \frac{1}{2}  \left[ \frac{\mathcal{\bar{A}}_+}{\mathcal{A}_+}  \pm 
      \left( \frac{\mathcal{A}_-}{\mathcal{\bar{A}}_-} \right)^* \right]. 
         \end{aligned}
   \end{equation}
  By applying symmetry operators to amplitude of zero-spin system, 
  relations between parameters introduced in Eq.~\ref{zabawne_wspolczynniki} and conservation of a
  particular symmetry~\cite{handbook_cp} can be obtained. Those relations are summarized in Table~\ref{table_fun1}.
    \begin{table}[h!]
    \centering
    \begin{tabular}{|c|c|} \hline
     Conserved quantity & Required relation \\ \hline \hline
     $\Delta S = \Delta Q$ rule & $x = \bar{x} = 0$ \\
     $\mathcal{CPT}$ symmetry & $x = \bar{x}^*$, $y=0$   \\
     $\mathcal{CP}$  symmetry & $x = \bar{x}$, $y =$ imaginary \\
     $\mathcal{T}$   symmetry & $y=$ real \\ \hline
    \end{tabular}
    \caption{Relations between discrete symmetries and semiletponic amplitudes}
    \label{table_fun1}
   \end{table}

		\noindent	
  Quantities from Eq.~\ref{zabawne_wspolczynniki} %  \ref{aplitudes_def}
  can be associated to the  $K_{S}$
  and $K_{L}$ semileptonic decay widths through the charge asymmetry~($A_{S,L}$):
   \begin{equation}
    \begin{aligned}
     A_{S,L} & = 
        \frac{\Gamma(K_{S,L} \rightarrow \pi^{-} e^{+} \nu) - \Gamma(K_{S,L}
        \rightarrow
        \pi^{+} e^{-}
        \bar{\nu})}{\Gamma(K_{S,L} \rightarrow \pi^{-} e^{+} \nu) + \Gamma(K_{S,L}
        \rightarrow
        \pi^{+} e^{-} \bar{\nu})}    \\
          & = 2 \left[  Re\left( \epsilon_{S,L} \right) - Re (y) \pm Re( x_{-}) \right].       
    \end{aligned}
   \end{equation}
  The above equation contains only the first order of  symmetry-conserving terms 
  with parameters $\epsilon_{S}$, $\epsilon_{L}$ %(\ref{ks_kl})
  which can be expressed in terms of
  the $\mathcal{CP}$ and $\mathcal{CPT}$ violation parameters $\epsilon_{K}$ and $\delta_{K}$. 

  Sum and difference of the $A_{S}$ and $A_{L}$ allow to search for the $\mathcal{CPT}$
  symmetry violation, either in 
  the decay amplitudes through the parameter $y$ (see~Table~\ref{table_fun1}) 
  or in the mass matrix through the parameter $\delta_{K}$: 
   \begin{equation}
    \begin{aligned}
     A_{S} + A_{L} &= 4 Re( \epsilon ) - 4 Re \left( y \right), \\
     A_{S} - A_{L}  %&= 4 Re( \delta_{K}) - 4 Re \left( \frac{d^{*}}{a} \right) %\nonumber \\
                    &= 4 Re( \delta_{K}) + 4 Re \left( x_{-} \right).
    \label{cpt_asymetria}
    \end{aligned}
    \end{equation}
  A precise measurement of the number of $K_S$ and $K_L$ semileptonic decays  allows to determine
  the value of charge asymmetry and tests $\mathcal{CPT}$ violation and $\Delta S = \Delta Q$ rule
  violation.
  The charge asymmetry for long- and short-lived kaons were determined by CPLEAR and KLOE
  experiments, respectively~\cite{kloe_final_semileptonic,CPLEAR_charge_asymm}. 
  Measurement of $A_L$ was based on $1.9$ millions $K_{L} \rightarrow \pi
  e \nu$ decays produced in collisions of proton beam with a BeO target.
  Following values were obtained~\cite{CPLEAR_charge_asymm}:
   \begin{equation}
    A_{L} = (3.322 \pm 0.058_{stat} \pm 0.047_{syst}) \times 10^{-3}.
   \end{equation}
  At present most accurate measurement of $A_S$ was performed with $0.41 \mbox{ fb}^{-1}$ total
  luminosity data sample and is equal~\cite{kloe_final_semileptonic}:
   \begin{equation}
    A_{S} = (1.5 \pm 9.6_{stat} \pm 2.9_{syst}) \times 10^{-3}.
   \end{equation}
   This result is consistent with the charge asymmetry determined for long-lived kaons within
			errors. 

\section{Measurement}
		Obtained results of $A_S$ and real part of $x_{+}$, $x_{-}$, $y$  parameters allow to perform the 
  most precise tests of  $\mathcal{CPT}$ symmetry and $\Delta S = \Delta Q$ rule  
  in semileptonic decays of neutral kaons. However, accuracy on $A_L$ determination is more than two
  orders of magnitude bigger than this of the $A_S$ and the uncertainty on  $A_S$ is dominated by
		the data sample statistics three times larger than the systematic contribution.
	\subsection{KLOE}
\begin{figure}[h!]
			\vspace{-10pt}
			\begin{center}	
		 \includegraphics[width=0.6\textwidth]{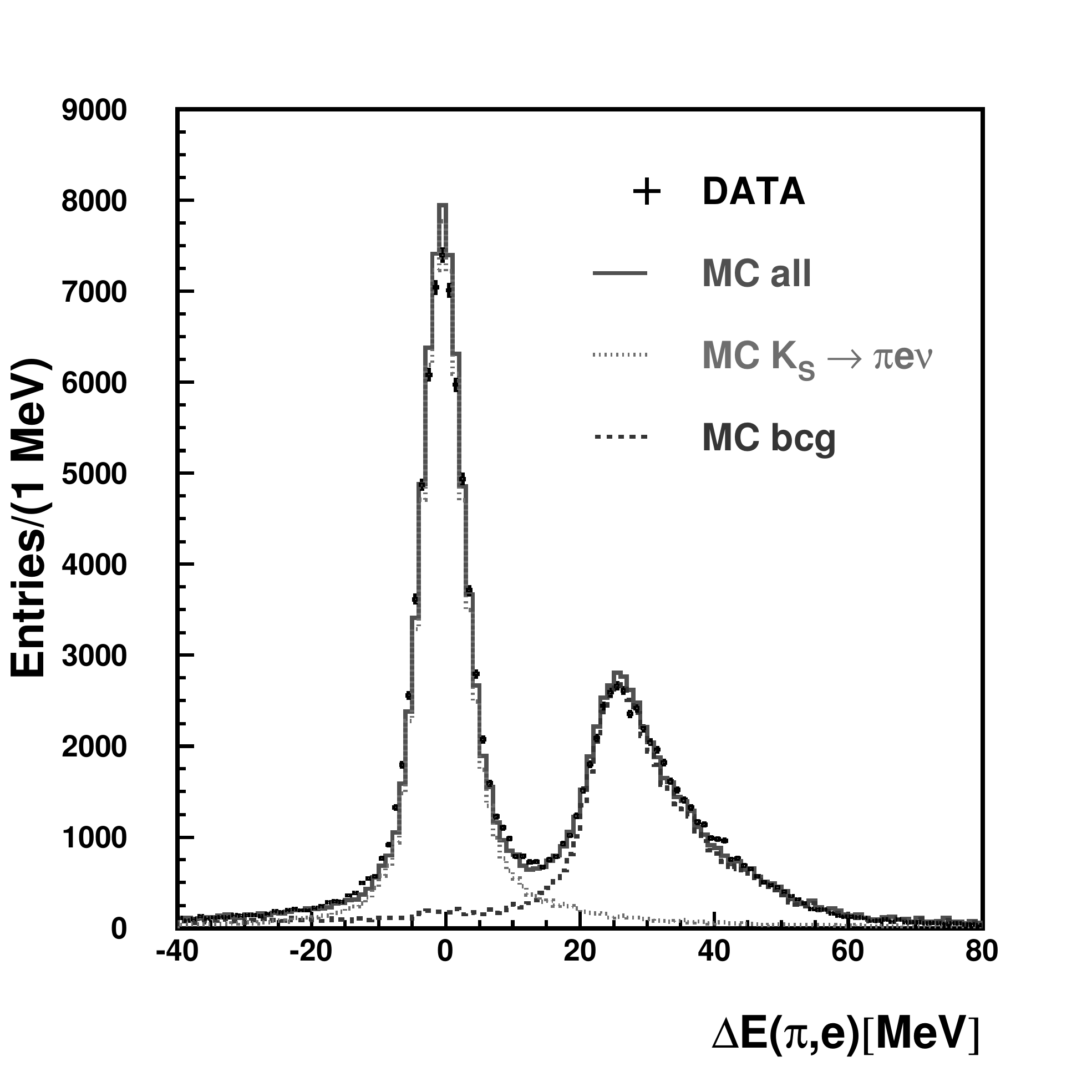}
		 \caption{Distribution of $\Delta E(\pi,e) = E_{miss} - p_{miss}$ for all selected events 
		 after normalization procedure.}
			\end{center}
			\vspace{-10pt}
		 \label{de_pi_e}
		\end{figure}

		The measurement is based on the ability to tag a $K_S$ meson by identifying  the $K_L$ meson.
		The KLOE detector consists of two main parts:
		a drift chamber~\cite{drift_chamber} and a barrel shaped electromagnetic
		calorimeter~\cite{calorimeter}, both inserted into a magnetic field
		(0.52~T). Around $60\%$ of $K_L$ mesons reach the electromagnetic calorimeter and can be identified 
		by their energy deposition inside~it. 
		The selection of  $K_S \rightarrow
		\pi e \nu$ decays requires a vertex reconstructed near the Interaction Point
		with two tracks that belong to two oppositely 
		charged particles.
		These particles must reach the
		calorimeter and deposit energy inside it in order to use
		Time of Flight  technique.  This technique aims at rejecting background, which
		consists mainly of  $K_S \rightarrow \pi \pi$ events, and at identifying the final charged states
		($\pi^+ e^- \bar{\nu}$ and $\pi^- e^+ \nu$). The distribution	of the difference between missing
		energy and momentum ($\Delta E(\pi,e)$) shows the
		remaining	background components (see Figure~\ref{de_pi_e}).
		Based on an integrated luminosity of $1.7 \mbox{fb}^{-1}$ 	around $10^5$ of $K_S \rightarrow \pi e \nu$ 
		decays were reconstructed and will be
		used to determine the charge asymmetry and branching ratio for $K_S$ semileptonic decays.
		A preliminary analysis shows a potential of reaching a two times better statistical error determination
		with	a sample four times bigger than the previous KLOE analysis.
		The analysis is still in progress and preliminary results will be 	available	soon.

		\newpage

	\subsection{Prospects for KLOE-2}		
		In the near future further improvements of both statistical and systematical uncertainty are expected
		thanks to the luminosity upgrade of DA$\Phi$NE and the installation of new sub-detectors in the KLOE-2
		experiment~\cite{prospects_kloe}. The improvement on kaon vertex reconstruction and acceptance for tracks with low transverse
		momentum in the region near the Interaction Point crucial for $K_S$ decays, will
		be ensured by  the newly installed Inner Tracker sub-detector~\cite{inner_tracker}.
		KLOE-2 is also equipped with low- and high- energy taggers that allow to identify $e^+ e^-$
		originated from $e^+ e^- \rightarrow e^+ e^- X$ reactions for $\gamma \gamma$
		physics~\cite{LET,HET}.
		Reconstruction of neutral particles
		at low polar angles will be improved due to the installation of CCALT~\cite{CCALT} and
		QCALT~\cite{QCALT} extra calorimeters. 
		It should be emphasised that KLOE-2 aims to significantly improve the sensitivity of tests of
		discrete symmetries, through studies of $K_S$ charge asymmetry or quantum interferometry effects in the
		kaon decays, beyond the presently achieved results~\cite{prospects_kloe,proposal}.

%\section{Acknowledgments}
\bigskip
		We warmly thank our former KLOE colleagues for the access to the data collected during the KLOE data taking campaign.
		We thank the DA$\Phi$NE team for their efforts in maintaining low background running conditions and their collaboration during all data taking. We want to thank our technical staff: 
		G.F. Fortugno and F. Sborzacchi for their dedication in ensuring efficient operation of the KLOE computing facilities; 
		M. Anelli for his continuous attention to the gas system and detector safety; 
		A. Balla, M. Gatta, G. Corradi and G. Papalino for electronics maintenance; 
		M. Santoni, G. Paoluzzi and R. Rosellini for general detector support; 
		C. Piscitelli for his help during major maintenance periods. 
		This work was supported in part by the EU Integrated Infrastructure Initiative Hadron Physics
		Project under contract number RII3-CT- 2004-506078; by the European Commission under the 7th
		Framework Programme through the `Research Infrastructures' action of the `Capacities' Programme,
		Call: FP7-INFRASTRUCTURES-2008-1, Grant Agreement No. 227431; by the Polish National Science
		Centre through the Grants No.  
		%0469/B/H03/2009/37, 
		%0309/B/H03/2011/40, 
		DEC-2011/03/N/ST2/02641, 
		2011/01/D/ST2/ 00748,
		2011/03/N/ST2/02652,
		2013/08/M/ST2/00323,
		and by the Foundation for Polish Science through the MPD programme and the project HOMING PLUS BIS/2011-4/3.


\begin{thebibliography}{99}
\bibitem{handbook_cp} L. Maiani, G. Pancheri, N. Paver,  INFN-LNF (1995).
\bibitem{kloe_final_semileptonic} KLOE Collaboration, {\it Phys. Lett.} {\bf B636}, 173 (2006).
\bibitem{CPLEAR_charge_asymm} CPLEAR Collaboration, {\it Phys. Lett.} {\bf B.444}, 52 (1998).
%\bibitem{ktev_kl_charge_asymm} KTeV Collaboration,  Phys. Rev. Lett. 88 (2002), p. 181601-181606
\bibitem{drift_chamber} KLOE Collaboration,{\it Nucl. Instrum. Meth.} {\bf A461}, 25 (2001).
\bibitem{calorimeter} KLOE Collaboration, {\it Nucl. Instrum. Meth.} {\bf A482}, 364 (2002).
\bibitem{prospects_kloe} KLOE-2 Collaboration, {\it Eur. Phys. J.} {\bf C68}, 619 (2010).
\bibitem{inner_tracker} G. Morello et al., {\it JINST} {\bf 9}, C01014 (2013).
% http://inspirehep.net/record/822246/export/hx
\bibitem{LET} KLOE Collaboration, {\it Nucl. Instrum. Meth.} {\bf A617}, 81 (2010).
\bibitem{HET} KLOE Collaboration, {\it Nucl. Instrum. Meth.} {\bf A617}, 266 (2010).
\bibitem{CCALT} M. Cordelli et al., {\it Nucl. Instrum. Meth.} {\bf A718}, 81 (2013).
\bibitem{QCALT} A. Balla et al., {\it Nucl. Instrum. Meth.} {\bf A718}, 95  (2013).
%proposal
\bibitem{proposal} KLOE-2 Collaboration,  LNF-10-17-P (2010).
\end{thebibliography}
\end{document}